\documentclass{article}

\usepackage{arxiv}

\usepackage[utf8]{inputenc} 
\usepackage[T1]{fontenc}    
\usepackage{hyperref}       
\usepackage{url}            
\usepackage{booktabs}       
\usepackage{amsfonts}       
\usepackage{nicefrac}       
\usepackage{microtype}      
\usepackage{lipsum}
\usepackage{graphicx}
\usepackage{graphicx}
\usepackage[utf8]{inputenc} 
\usepackage[T1]{fontenc}    
\usepackage{hyperref}       
\usepackage{url}            
\usepackage{booktabs}       
\usepackage{amsfonts}       
\usepackage{nicefrac}       
\usepackage{microtype}      
\usepackage{lipsum}		
\usepackage{slashbox}
\usepackage{lscape}
\usepackage{subcaption}
\usepackage{placeins}
\graphicspath{ {./images/} }

\title{Presenting a Dataset for Collaborator Recommending Systems in Academic Social Network: a Case Study on ReseachGate}

\author{
 Zahra Roozbahani \\
  Department of computer Engineering and IT\\
  University of Qom,\\
  Qom, Iran \\
  \texttt{z.roozbahani@stu.qom.ac.ir} \\
   \And
 Jalal Rezaeenour  \\
  Department of Industrial Engineerin\\
  University of Qom,\\
  Qom, Iran \\
  \texttt{j.rezaee@qom.ac.ir} \\
  \And
 Roshan Shahrooei \\
  Department of Industrial Engineerin\\
  University of Qom,\\
  Qom, Iran \\
  \texttt{Shahrooei@yahoo.com} \\
  \And
  Hanif Emamgholizadeh \\
  Department of computer Science\\
  Yazd University\\
  Yazd, Iran\\
  \texttt{h.emamgholizadeh@gmail.com} \\
}

\begin{document}
\maketitle
\begin{abstract}
Collaborator finding systems are a special type of expert finding models. There is a long-lasting challenge for research in the collaborator recommending research area, which is the lack of a structured dataset to be used by the researchers. We introduce two datasets to fill this gap. The first dataset is prepared for designing a consistent, collaborator finding system. The next one, called a co-author finding model, models an academic, social network as a table that contains different relations between the pair of users. Both of them provide an opportunity for introducing potential collaborators to each other. These two models have been extracted from ResearchGate (RG) data set and are available publicly. RG dataset has been collected from Jan. 2019 to April 2019 and includes raw data of 3980 RG users. The dataset consists of almost complete information about users. In the preprocessing phase, the well-known Elmo was used for analyzing textual data. We call this as \textbf{R}esearch\textbf{G}ate dataset for \textbf{R}commending \textbf{S}ystems (\textbf{RGRS}). For assessing the validity of data, we analyze each layer of data separately, and the results are reported. After preparing data and evaluating the collaborator finding models, we have done some assessments on \textbf{RGRS}. Some of these assessments are co-author, following-follower, and question answering relations. The outcomes indicate that it is the best relation in propagating knowledge in the network. To the best of our knowledge, there is no processed and analyzed dataset with this size.
\end{abstract}


\section{Introduction}
Recommender systems are a subset of decision support systems. Recommender system is a collaborator recommender system in scientific environments. Scientific collaboration can be defined as a behavior among two or more scientists, which facilitates the description of the task and the positions for each member in collaboration concerning a predefined objective \cite{sonnenwald2007scientific}.\\

In the field of scientific collaborator finding systems, researchers have provided different models that have been completely introduced and studied in the article \cite{Roozbehani2020survey}.\\

One of the most important concerns of researchers in finding a collaborator and designing collaborator recommender systems is the lack of suitable structured data to examine collaborator finding models. Many datasets of scientific social networks have been published so far. However, there is no exhaustive system for recommending scientific collaborators. Researchers generally use scientific social networks to find potential collaborators. Indeed, scientific and academic social networks, such as mendely\footnote{https://www.mendeley.com/}, Academia\footnote{https://www.academia.edu/}, Aminer\footnote{https://aminer.org/}, LinkedIn\footnote{https://www.linkedin.com/}, and ResearchGate\footnote{https://www.researchgate.net/} provide some facilities to be utilized by the users for enhancing their knowledge about other users in the networks and finding new collaborators for their current or future projects. The existing data,extracted from these social networks, can be divided into two main groups structural information and personal information. Following-Follower information, collaboration relations, etc. are called structural information that stems from the attributes and abilities of the social networks for representing such information. Personal information, on the other hand, is specific for each user and includes affiliation, department, interests, and so forth. These types of information have nothing to do with the network capabilities and only demonstrate the users' personal information.\\

Several public data sets have been extracted and published about scientific social networks. Still, some shortcomings make new and comprehensive data set of scientific social networks an immediate need. One of the most obvious drawbacks of published data set is considering only one of the two main information sources, structural or personal. Another protrude defect is the lack of a meaningful frame for the collaborator recommendation problem.\\

For instance, different datasets of DBLP have been published so far \cite{davoodi2012social,hoang2017consensus}, but this data does not include the structural information of the communication network between individuals and only includes the personal information of the researchers. Another dataset provided in the field of collaborator finding is the ResearchGate database \cite{liang2018modeling}, which includes only information on user articles. Other datasets have been provided in this field, many of which have not been made available to the public.\\

Our main objective in this research is to provide a suitable data structure for assessing scientific collaborator recommender systems. We attempted to prepare a comprehensive dataset of useful and practical information for collaborator selection. Given that among the existing scientific social networks, the ResearchGate network is the most extensive scientific network with the most comprehensive dimensions of personal and scientific information of users; user information has been collected from this network. In the first step, all information related to 3980 users of the ResearchGate social network in the period from January 2019 to April 2019 was collected. In the second step, data cleaning and preparation operations were performed. In this step, processes such as content analysis and structure analysis were performed. One of the most important algorithms used in content analysis was the elmo algorithm. In the third step, the raw data were prepared in the form of two separate models for assessing the collaborator finding models. In the first model, the scientific social network is modeled as a multi-relational directed network. This data structure can be used to analyze collaboration relationships and design a comprehensive collaborator recommender system. The second proposed structure is the modeling of the scientific social network in the form of a pairwise comparison table of users, which allows users to introduce co-authors. This model is also useful for assessing recommender systems.This data is publicly available for researchers\footnote{https://github.com/zroozbahani/RGData}.

The rest of the present paper is organized as follows. In the Sec. \ref{Literature_review}, we review the related researches. Then, in Sec. \ref{Raw_Data}, we describe and analyze our data. Afterward, in Sec. \ref{evaluation}, we evaluate the presented data set by a number of methods. Finally Sec. \ref{Conclusion}, conclude this paper.

\section{The Literature Review}
\label{Literature_review}
As mentioned before, there are several data set for evaluating the collaborator recommending models, which some of them have been published and are freely available. In the \cite{Roozbehani2020survey} the presented datasets that are used in recommending collaborator in scientific social networks' papers have been discussed comprehensively. In the rest of this section, we briefly describe some of these data sets.\\
\begin{itemize}
	\item \textbf{ArentMiner}\footnote{http://uvt-expert-collection/ }\cite{sujatha2012similar}: This data set has been used for finding similar researcher with respect to a query. The data have been collected from CiteSeerX documents, a website that includes researches information. The including data are information about the users who participated in a research project. This data set consists of 16 questionnaires, research field of publications, research website, home pages of users, and users' pages. This data set has been extracted in 2012 and includes information of 15000 users, and is freely available. The data set solely contains personal data and publication information, which confines the precision of collaborator models.\\
	\item \textbf{Research University \cite{anongnart2012building}: } This data set has been collected from Thailand national university and includes profile information of university professors. The data is about the project and consists of several sections. Research interests' section includes these data: ID number, first and last name, birth date, academic degree, job position, and the beginning date of work. Alumna data section consists of ID number, Adviser name, the title of the thesis, and period of study. Finally, the project section data contains ID number, researcher name, and research title. This data set has been gathered in 2012 and includes 3194 users. This data set is not available publicly. The data set is a bit more comprehensive than the aforementioned data sets; however, this comprehensiveness returns to personal information, whereas there is no structural information.
	\item \textbf{DBLP}\footnote{http://www.informatik.uni-trier.de/ ley/db}\cite{davoodi2012social}: This data set consists of a list of journals used by each researcher. Information on this data set has been collected by a list of keywords and information from scientific journals, like google scholar. The included fields are the research field, scientific experiences, abilities, and social and political activities. The data set has been extracted in 2012 and includes information of 315 researchers and 62 research topics, and is publicly available for researchers. The data set is publicly available.
	\item \textbf{Academic researcher network \cite{xu2012combining}: }The information has been collected from the Asia-Australia annual meeting and its attributes: name of authors and information of publications' keywords. The data set has been collected from 2007 to 2009. The data set is not freely available.
	\item \textbf{Microsoft Academic Search website}\footnote{http://academic.research.microsoft.com/} \cite{huynh2014collaborator}: This data set has been collected using Microsoft search engine, and the fields consist of the name of authors and keywords of publications. Dataset has been collected from 2001 to 2011 and includes 1266790 journals and 807005 authors. The data set is publicly available for researchers. The information about the users is extremely small; therefore, it provides results with low accuracy.
	\item \textbf{Google Scholar: \cite{rani2015expert}} This data set has been gathered from google scholar, a search engine for scientific publications. The data has been extracted from 2013 to 2014 and includes 30 journals. Dataset is not available publicly, and there is no more information about the attributes and information in the data set.
	\item \textbf{DBLP-2}\footnote{http://dblp.uni-trier.de/} \cite{hoang2017consensus}: This data set has two main characteristics: (i) it includes more than 3.4 million journals, more than 1.7 million authors, and 4896 conferences, which made it popular to be used by the researchers, (ii) all the references have been included in this data set. The containing information is the title of the paper, the name of the author, date of publication, type of publication, the title of conferences, and DOI links. Data have been collected in 2016 and is publicly available.  This data only consists of structural information, and the corresponding recommendation is just able to recommend collaboration based on prior collaborations. In addition, this data set is more proper for the content analyzer to analyze the content information data presented in papers. It seems that for filling this gap, a new data set containing both structural and personal data is necessary.
	\item \textbf{ResearchGate}\footnote{https://archive.org/details/stackexchange} \cite{liang2018modeling}: One of the biggest scientific online social networks is ResearchGate. In this network, users are able to share their projects and publications. The collected data from this network includes 9000 researchers and 350000 publications published from 1986 to 2016. This data set is publicly available. Although the data set is big enough, there is not a sufficient amount of personal information, and this makes the recommendation results less accurate.
	\item \textbf{DBLP-3} \cite{liang2018modeling}: This data set has been collected from DBLP and includes a list of keywords, abstracts of digital libraries, and Google Scholar. The presented fields in this data set include the title of publication, author name, publisher name, date of publication, and DIO link. The data set has been made of 5000 randomly selected publication that published from 1986 to 2016. The data set is not publicly available, which is the main shortcoming of this data set, but not the only one. The data set also has the mentioned drawbacks of DBLP-2 and WikiCFP.
	\item \textbf{Mendeley} \cite{li2013identifying}: Mendeley is a social network in which the users can make their own profiles, add some information about themselves, upload their publications, and make their own research groups. The fields in this data set are common among direct users, members, and collaborators. These features include research interests, academic status, biographic information, publication title, abstract, conference, and number of readers. The data has been collected in 2012 and includes 1 million users and 10 million papers. This data set is not freely available; however, it contains the most comprehensive information about the authors; in the same breath, there is no structural information, such as follower, following,  personal skills, etc.
\end{itemize}
We have discussed the presented data for collaborator finding above; in addition, the collaborator finding researchers also can look for new and potential collaborations in social networks. To do so, the researchers should use provided facilities by online social networks. Therefore, we will describe the presented facilities by the social networks for aiding the users in spotting potential collaborators. Table \ref{table1} indicates the presented abilities by the social network; this table shows what kind of information can be extracted in each social network. The variation of presented information is important for making a collaborator recommendation model with high quality; nevertheless, the social network should inherently be able to provide these kinds of information. If the developed model needs more information, which does not exist in the social network, the model would not have its best performance. Thus, the presented information and features in each social network should be taken into account before evaluating the model, and the fittest data of social networks should be considered as the evaluation data set.

\begin{table*}
	\centering
	\caption{The Extracted Information from each social network \cite{roozbahani2018}}
	\resizebox{1\textwidth}{!}{%
		\begin{tabular}{ |c | c | c| c|c | c |}
			\hline	
			\backslashbox{Features}{Social networks}& Academia & Mendely & LinkedIn  & Aminer & ResearchGate \\ \hline	
			Following & +  & + & +  & + & + \\ \hline	
			University &  & & +  & & + \\ \hline	
			Ethnicity and nationality &  & &   + & &  + \\ \hline	
			Collaboration&  & + &  + &  + & + \\ \hline	
			Major&  & + &  + &  + & +\\ \hline	
			Number of publication& +  & &  & +  & + \\ \hline	
			Question and answer relation& +  & private  & private & private & + \\ \hline	
			Tendency to accept collaboration suggestion&  & & + & &\\ \hline	
			Activity in social network&  & &  & & + \\ \hline	
			Scientific level&  & + & + & + & + \\ \hline	
			Participating in same conferences&  & + & + & + &\\ \hline	
			Co-citing by published paper&  & &  & + & + \\ \hline	
			Number of publication reader& + & + &  & & + \\ \hline	
			Number of profile visitors& + & &  + & & + \\ \hline	
			Number of users who have visited both candidates profiles&  & & + & & + \\ \hline	
			Relation in other social networks& + & + & +  & &\\ \hline	
			Following other collaborator's collaborators & + & + & +  & & + \\ \hline	
			Research interests& +  & + &  & + & + \\ \hline	
			\hline	
			Direct collaborator suggestion ability & \textbf{limited}  & \textbf{limited} & \textbf{limited} & \textbf{limited} & \textbf{limited} \\ \hline	
		\end{tabular}
	}
	\label{table1}
\end{table*}

Different studies have been carried out in the field of scientific collaborator finding. Some of these studies have identified researchers who are expected to collaborate in the future by co-authorship network analysis as well as using link prediction algorithms. \cite{sun2019career,torkzadeh2018expert}. Considering that the researcher's recommendation is an example of expert recommendation in scientific fields. Studies have provided expert finding models in scientific communities, \cite{davoodi2012social,sohangir2018finding} In these studies, it has been pointed out that the expert finding model can be used as a collaborator recommender system. Other researchers have attempted to identify researchers whose profiles are similar and introduce them as collaborators \cite{gollapalli2012similar}. In these studies, content analysis and network analysis methods have been used to find similar individuals.\\

In the studies conducted, various data have been used by researchers to assess collaborator finding models, some of which have been made available to the public. In the research (our review article), the collaborator finding models and the data used in them are completely provided. In the following, some of these data will be reviewed.

\section{Data}
\label{Raw_Data}
As the first step of providing a dataset, in this section, we describe how the raw data collected and processed.
\subsection{Raw Data}
Up to now, we have reviewed the presented data extracted from scientific social networks. As has been shown, there are some shortcomings like incomprehensiveness; moreover, most of the data are not publicly available. Hence, there is an immediate need for a freely available comprehensive data set to pave the way for researches in the related research field. In the rest of this section, we introduce the raw data which we have extracted from ResearchGate.\\

To collect the raw data, we developed a crawler using python. This crawler collected available data of users using a snowball method. Generally, this data set includes 13 tables. Each of these tables and their rows is shown in Table \ref{table2}. In the following, we introduce each of these tables in detail.

\begin{table}
	\centering
	\caption{Name and Type of Data}
	\resizebox{.5\columnwidth}{!}{%
		\begin{tabular}{ |c | c | }
			\hline	
			Type & 	Number of rows \\ \hline
			Users & 	3980 \\ \hline
			Answers & 	3426 \\ \hline
			Current researches & 	5178 \\ \hline
			Awards & 	1057 \\ \hline
			Followers & 	309698 \\ \hline
			Following & 	381974 \\ \hline
			Researchers interests & 	47340 \\ \hline
			Research interests & 	19725 \\ \hline
			References & 	123477 \\ \hline
			Research experience & 	4195 \\ \hline
			Authors & 	697481 \\ \hline
			Research items & 	225760 \\ \hline
			Skills & 	Number of rows \\ \hline
		\end{tabular}
	}
	\label{table2}
\end{table}

\textbf{Table of users:} Users' table includes RG users' information and contains information of 3980 members of RG Table \ref{table3}. In the users' table, there are 18 features (columns) for each user. Concerning this information, different aspects of users can be detected. Having information on the users makes this table the most important table, and the information on other tables has been collected based on this table. 
\begin{table*}
	\centering
	\caption{Users' Information}
	\resizebox{1\textwidth}{!}{%
		\begin{tabular}{ |c | c |c | c| }
			\hline	
			Column &  Column name	&Description & Data type \\ \hline
			1 & 	Username & User name of account & Textual \\ \hline
			2 & 	Name & Name of user & Textual \\ \hline
			3 & 	Score & RG score & Numerical \\ \hline
			4 & 	Education & Academic status of user& Textual \\ \hline
			5 & 	Research\_items & Number of publications and research items& Textual \\ \hline
			6 & 	User\_Reads & Number of readers for user's publications& Numerical \\ \hline
			7 & 	Citations & Number of citation for user's publications& Numerical \\ \hline
			8 & 	break\_publications& RGscore for publications & Textual \\ \hline
			9 & 	break\_questions& RGscore for questions & Textual \\ \hline
			10 & 	break\_answers& RGscore for answers & Textual \\ \hline
			11 & 	break\_followers& RGscore for followers & Textual \\ \hline
			12 & 	Percentile & proportion of user's score to average score of all members& Numerical \\ \hline
			13 & 	H\_index\_1 & is equal to $x$ if $x$ publications have at least $x$ citations&  Numerical \\ \hline
			14 & 	H\_index\_1 & is equal to $x$ if $x$ publications have at least $x$ citations regardless of self citation &Numerical \\ \hline
			15 & 	Questions& Number of asked questions & Numerical \\ \hline
			16 & 	Answers &Number of answers & Numerical \\ \hline
		\end{tabular}
	}
	\label{table3}
\end{table*}

\textbf{Table of answers:} The information of asked and answer questions is presented in Table \ref{table4}. This table shows the ID of the users who asked or answered the questions.\\
\begin{table}[h]
	\centering
	\caption{Question and Answers}
	\resizebox{.5\columnwidth}{!}{%
		\begin{tabular}{ |c | c | c| }
			\hline	
			Column & 	Description & Data type \\ \hline
			1 & 	Data ID & Numerical \\ \hline
			2 & 	ID of user who answered & Numerical \\ \hline
			3 & 	ID of user who asked the question & Numerical \\ \hline
		\end{tabular}
	}
	\label{table4}
\end{table}

\textbf{Table of followers:} Table \ref{table5} provides information of followers and following. This table includes two columns. The first column contains the users who have followed the users in the second column. Each user is shown by an ID, which can be used to find the corresponding user's information in other tables.\\
\begin{table}[h]
	\centering
	\caption{Following}
	\resizebox{.5\columnwidth}{!}{%
		\begin{tabular}{ |c | c | c| }
			\hline	
			Column & 	Description & Data type \\ \hline
			1 & 	Data ID & Numerical \\ \hline
			2 & 	Following's ID & Numerical \\ \hline
			3 & 	Follower's ID& Numerical \\ \hline
		\end{tabular}
	}
	\label{table5}
\end{table}

\textbf{Table of following:} Table \ref{table6} represents the table of followings. This table includes two columns. The users in the first column are followed by the users in the second column. Each user are indicated by an ID to be used for extracting information of him or his in other tables.\\

\begin{table}[h]
	\centering
	\caption{Followers}
	\resizebox{.5\columnwidth}{!}{%
		\begin{tabular}{ |c | c | c| }
			\hline	
			Column & 	Description & Data type \\ \hline
			1 & 	Data ID & Numerical \\ \hline
			2 & 	Follower's ID & Numerical \\ \hline
			3 & 	Following's ID& Numerical \\ \hline
		\end{tabular}
	}
	\label{table6}
\end{table}

\textbf{Table of skills:} Table \ref{table7} shows the information of users' skills. This table contains two columns the first one is users ID and the second columns includes the users skills.\\
\begin{table}[h]
	\centering
	\caption{Skills and expertise}
	\resizebox{.5\columnwidth}{!}{%
		\begin{tabular}{ |c | c | c| }
			\hline	
			Column & 	Description & Data type \\ \hline
			1 & 	Data ID & Numerical \\ \hline
			2 & 	Author ID & Numerical \\ \hline
			3 & 	User expertise& Numerical \\ \hline
		\end{tabular}
	}
	\label{table7}
\end{table}

\textbf{Table of published paper:} In table \ref{table9}, we introduce the information about publications of users. As the users' table, all of the publications of these users have been inserted in the table with a unique ID.

\begin{table}[h]
	\centering
	\caption{Publications}
	\resizebox{.5\columnwidth}{!}{%
		\begin{tabular}{ |c | c | c| }
			\hline	
			Column & 	Description & Data type \\ \hline
			1 & 	Publication ID & Numerical \\ \hline
			2 & 	User ID & Numerical \\ \hline
			3 & 	Publication title& Textual \\ \hline
			4 & 	Publication link & Textual \\ \hline
			5 & 	Number of readers & Numerical \\ \hline
			6 & 	Number of citation& Numerical \\ \hline
			7 & 	Number of citation & Numerical\\ \hline
			8 & 	Title of publication & Textual \\ \hline
			9 & 	Type of gained information & Textual \\ \hline
			10 & 	Location & Textual \\ \hline
			11 & 	Published date & Textual \\ \hline
			12 & 	Publication status& Textual \\ \hline
			13 & 	Abstract & Textual \\ \hline
			14 & 	Journal or conference name& Textual \\ \hline
		\end{tabular}
	}
	\label{table9}
\end{table}

\textbf{Table of authors:} Table \ref{table8} demonstrates information of authors of the publication. The number of each publication has been extracted from publication of Table \ref{table9}.\\
\begin{table}[h]
	\centering
	\caption{Publication authors}
	\resizebox{.5\columnwidth}{!}{%
		\begin{tabular}{ |c | c | c| }
			\hline	
			Column & 	Description & Data type \\ \hline
			1 & 	Data ID & Numerical \\ \hline
			2 & 	Publication ID & Numerical \\ \hline
			3 & 	Author name& Textual \\ \hline
		\end{tabular}
	}
	\label{table8}
\end{table}


\subsection{Preprocessing}
Having collected the raw data, we have to do some processes on it to prepare \textbf{RGRS}. As mentioned before, there are two sources of data: structural and personal. Structural information, such as following, follower, or co-authoring, does not need any further processing to be used; however, personal data needs to be processed to be prepared. In the rest of this section, we explain how we prepared data.\\


\textbf{Ranking features:} For calculating H-Index and RG rank for pair users, we have used relation \ref{eq1}. \cite{cai2005mining}. This relation provides a value between 0 and 1 because the range of $\exp$ function is from zero to an infinite number. The negative element in the power of $\exp$ caused the smallest distance between two users to be 0 (when two users have the very same H-index), while the biggest distance is reflected by 1 when the H-index distance between the users is an infinite number.

\begin{equation}
	\label{eq1}
	y = \exp^{-\{|x_1 - x_2|\}}
\end{equation}

\textbf{Textual features:} The aforementioned method is not applicable for textual data, such as department name, research similarity, and skill similarity. These features should be transformed into numerical values. This transformation was done by Elmo. Elmo presented by Peters et al. \cite{peters2018deep}, which has been claimed to have 84.6 accuracy on textual data. This algorithm is an embedding algorithm to transform textual representation to vector representation. Textual data of this data set has been transformed into the vectors by this algorithm, and cosine similarity has been used for calculating the distance between two vectors.\\

For processing textual data, first, the abstracts of publications of each user have been extracted using mean pooling, and the sentences of these textual data has been transformed into a vector. Finally, the cosine similarity of the publication for each publication of each user with each publication of another user has been calculated, and a matrix for each pair of users has been made. The value of features for publication similarity for each of two users has been assigned as a mean value of this similarity matrix. For other textual values, the same process has been applied as well. For example, the name and description of the department for each user transformed into a vector using Elmo, and these vectors cosine similarity inserted to the corresponding cell in the table.\\

Skill similarity of each pair of users has been calculated based on the portion of overlapped skills between two researchers. This value is fitted for each user; in other words, the value of skill similarity between $u_a$ and $u_b$ is not necessarily the same. This difference originates from the denominator of the measure, which is not the same among the users. For example, $u_a$ has 5 skills which two of them are the same as $u_b$ who has 10 skills. Thus, similarity of $u_a$ with $u_b$ is $\frac{2}{5}$, while the similarity of $u_b$ and $u_a$ is $\frac{2}{10}$.\\

\subsection{Dataset analysis}

In order to have a better understanding of the dataset, we analyze follower, following, and QA relations using community detection algorithm.\\

\begin{figure}
	\centering
	\includegraphics[width=0.8\columnwidth]{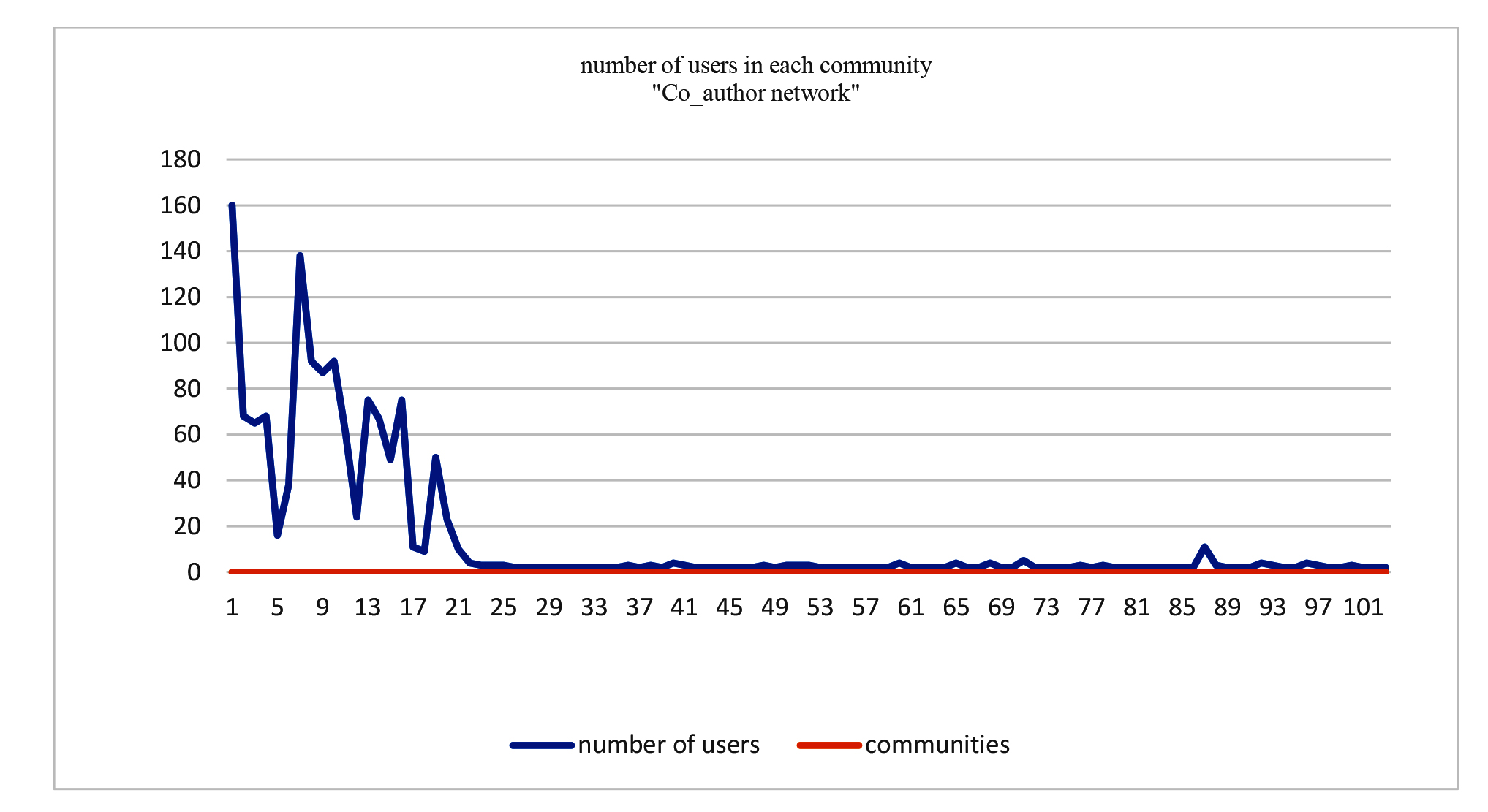}
	\caption{Figure of co-authoring communities}
	\label{Co-author_community_fig}
\end{figure}

\textbf{Analyzing communities in co-authoring network:} The co-authoring network contains 103 communities in which the community with the largest number of members includes 160 users, and the community with the smallest number of members has only 2 users. Fig. \ref{Co-author_community_fig} indicates the figure of communities in the co-authoring network. The average number of members in communities is 14.4, and the modularity of communities is 0.7918, which shows the detected community are well-separated. As is evident in Figure \ref{Co-author_community_fig}, clusters of co-authoring networks are small and mostly mutually exclusive. This indicates that researchers have collaboration with specific users, and also, there is a small amount of collaboration.\\

\begin{figure}
	\centering
	\includegraphics[width=0.8\columnwidth]{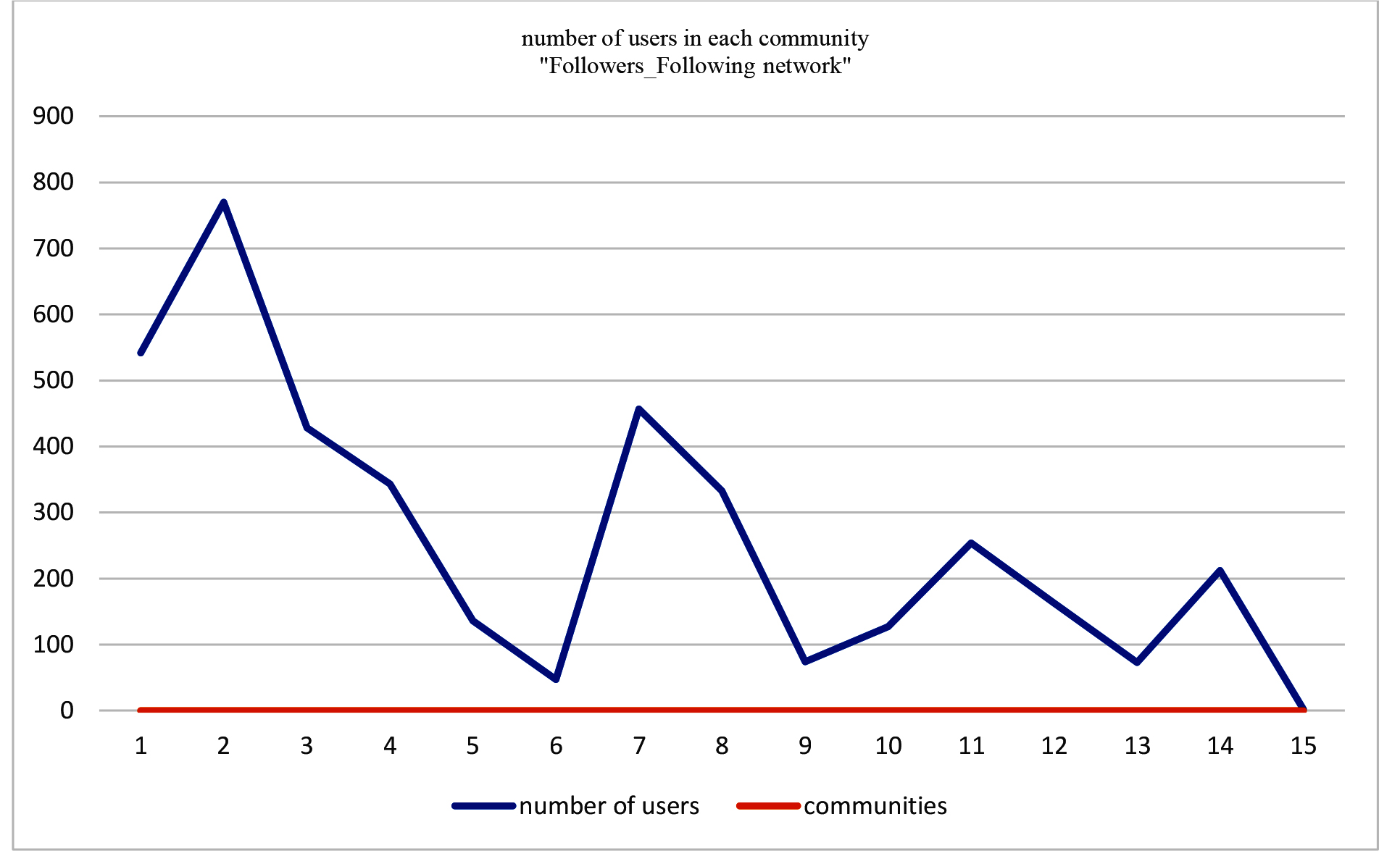}
	\caption{Figure of follower/following communities}
	\label{follower_community_fig}
\end{figure}

\textbf{Analyzing communities in follower/following network:}Number of communities of follower/following network is 15 communities in which the community with the largest number of member contains 770 users, and the community with the smallest number of the community includes only 2 users (Figure \ref{follower_community_fig}). The average number of users in communities is 264.06, and the modularity of communities 0.5168, which means the detected communities are meaningful enough. The network has a high density for detected communities. The users are divided into communities with high overlap. They are divided into 15 clusters with 250000 users, on average. This network has a higher density compared to co-authoring and question-answer networks.

\begin{figure}
	\centering
	\includegraphics[width=0.8\columnwidth]{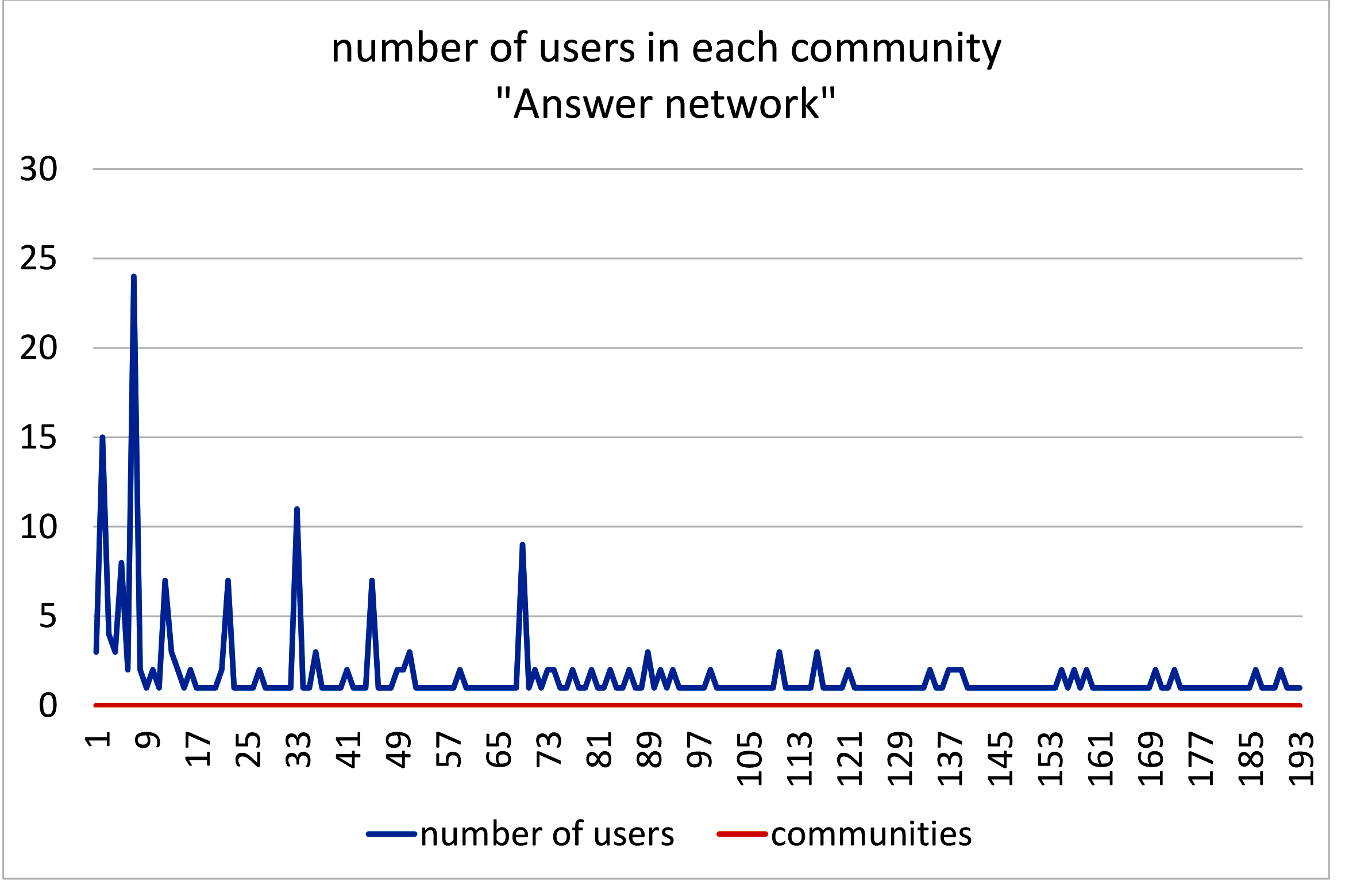}
	\caption{Figure of question-answer communities}
	\label{QA_community_fig}
\end{figure}

\textbf{Analyzing communities in Question Answer network:} There are 193 communities in question and answer networks, in which the community with the biggest number of users contains 24 users and the community with the smallest number of users includes only one user. The detected communities have 0.9736 modularity, which shows a high quality of detected communities. As it is evident from Figure \ref{QA_community_fig} and the detected communities in the question-answer network, there is a lot of small clusters. This network is similar to the co-authoring network, but networks overlap smaller in this network. This indicates question-answer networks have a low density in which users have fewer relations.

By analyzing hidden communities in co-authoring, follower/following, and QA networks, we want to determine which of these relations makes more meaningful communities and contain more detectable knowledge since including detectable knowledge is an important feature of social networks; therefore, we want to explore the most informative relation.\\

Clustering coefficient in Table \ref{table11} demonstrates that users are not interested in making communities using QA network; however, they make strong and meaningful communities based on follower and following network, and co-authoring relation, and also makes meaningful communities, with a bit less convergence in communities compared to follower and following network.\\

In Table \ref{table11}, it is clear that the QA network makes the most number of communities with the least average number of users in each community.  This shows although the QA network makes a great number of communities, the communities are not able to attract the attention of quite a few users. Therefore, the produced knowledge in each community diffuses among only 2 users. Indeed, follower/following network shares the produced information among a great portion of network users because there are a few communities with a great number of users (on average 264 users in each community).\\

Finally, with respect to Table \ref{table11} co-authoring network are weaker relation in comparison to follower/following and stronger compared to QA relation. Notwithstanding, co-authoring is a meaningful relationship since there is a great number of communities (collaboration seeds) with many users in each community (on average 14) which can collaborate, directly or indirectly, on a project with one another.\\

In summary, we can claim that based on our studies, QA relations attract less attention to users; hence, it is less meaningful in making new relations. Because collaborations transmit information in the network, QA relations are not able to play a pivotal role in this regard. On the other hand, follower/following relation seems to be a paved road for diffusing knowledge among the users. In other words, by establishing new links (follower/following) among the users, we can make knowledge diffusion much easier.

\begin{table*}
	\centering
	\caption{Information of Co-authoring, Follower/Following, and QA Networks \& Their Communities. CC stands for clustering coefficient; MaxNU stands for Maximum number of users; MinNu stands for Minimum number of users; ANU stands for Average number of users.}
	\resizebox{1\textwidth}{!}{%
		\begin{tabular}{ |c | c | c|c |c | c | c| c |}
			\hline	
			& CC& \#Nodes & \#Edges& 	Communities & MaxNU & MiniNU & ANU\\ \hline
			Co-authoring& 0.37 & 143& 	3413 & 103 & 160 & 2 & 14.4\\ \hline
			Follower.Following& 0.34& 396 & 33489& 	15 & 770& 2&264.06 \\ \hline
			Question/Answer& 32 & 354& 0.0 & 	193 & 24 & 1&1.68 \\ \hline
		\end{tabular}
	}
	\label{table11}
\end{table*}
\section{Evaluation}
\label{evaluation}
For the evaluation of our datasets in this section, we examine them using two collaborator recommendation models; thereby, the functionality of the datasets is tested. In the rest of this section, these evaluations are described.
\subsection{Consistent Collaboration Model}
By considering collaborator finding methods, there is a clear tendency toward combined models \cite{sohangir2018finding}. The combined models combine structural and content information. There are two challenges in this regard, the first is content analysis, and another one is the combination of the results of these two. In the considered models, structural and content information are evaluated separately, and two kinds of ranking are provided. The final ranking is calculated using these two completely separated rankings. For keeping the consistency of the model, we propose to analyze structural and content information at the same time. One of the methods is using multilayer networks. The only research which has used this idea is \cite{xu2012combining}. In this research, the author has used a two-layer network, one for users, another one for concepts. But, we propose using the maximum amount of available information consistently in a multilayer network, which we believe will improve the quality of suggested collaborations.\\

\begin{figure*}
	\centering
	\begin{subfigure}[t]{0.45\textwidth}
		\includegraphics[width=\textwidth]{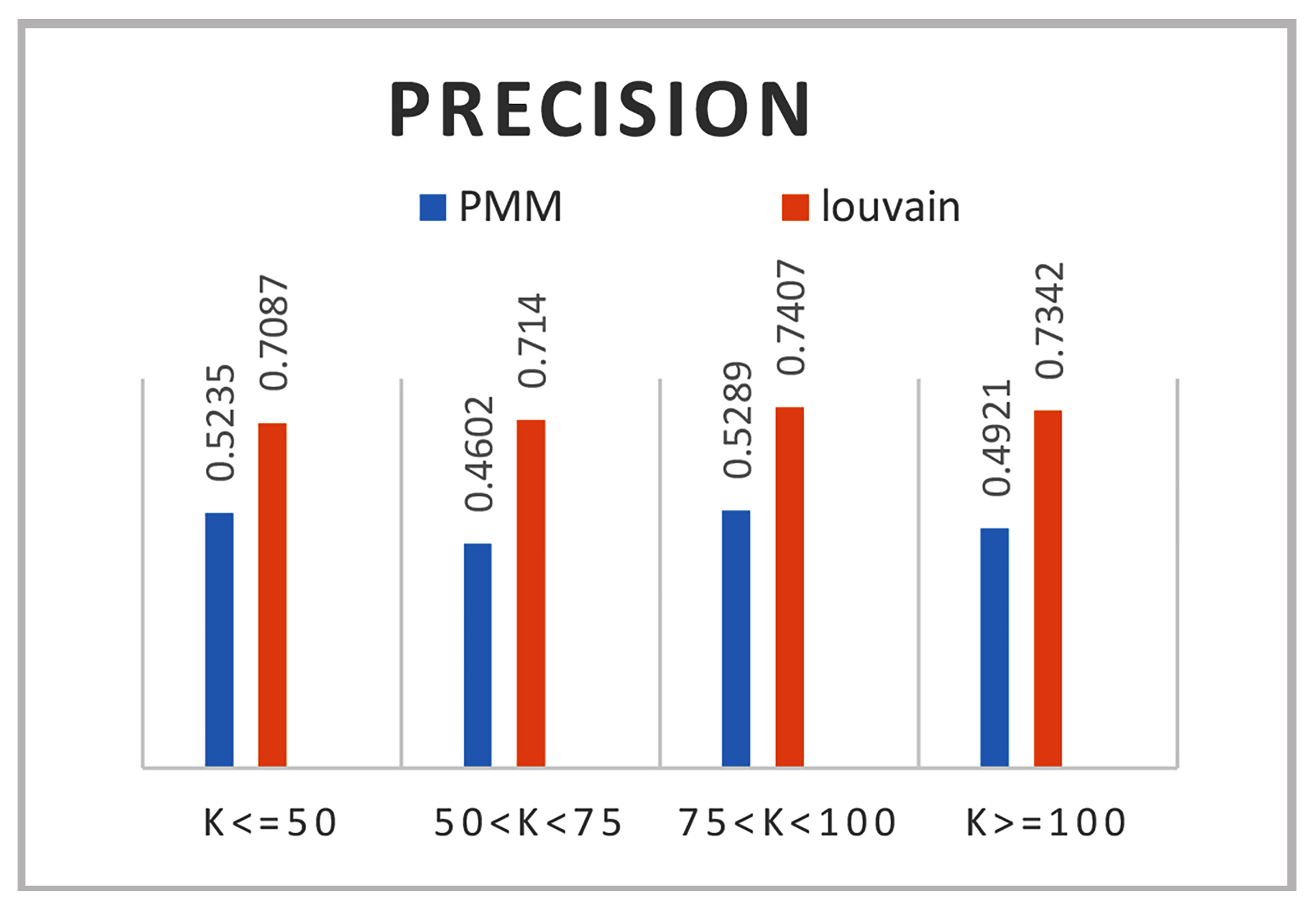}
		\label{prec}
	\end{subfigure}
	\begin{subfigure}[t]{0.45\textwidth}
		\includegraphics[width=\textwidth]{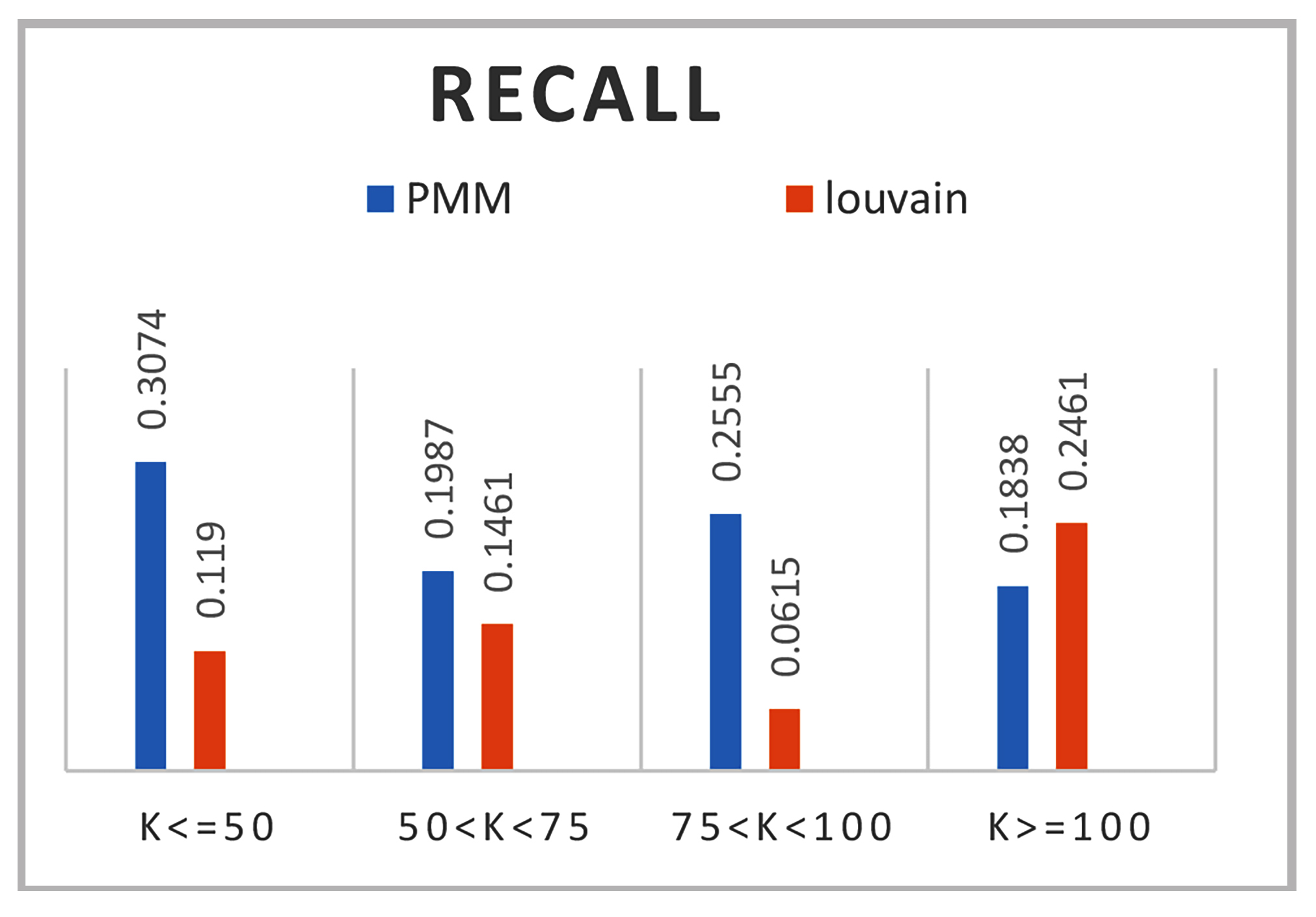}
		\label{recall}
	\end{subfigure}
	\begin{subfigure}[t]{0.45\textwidth}
		\includegraphics[width=\textwidth]{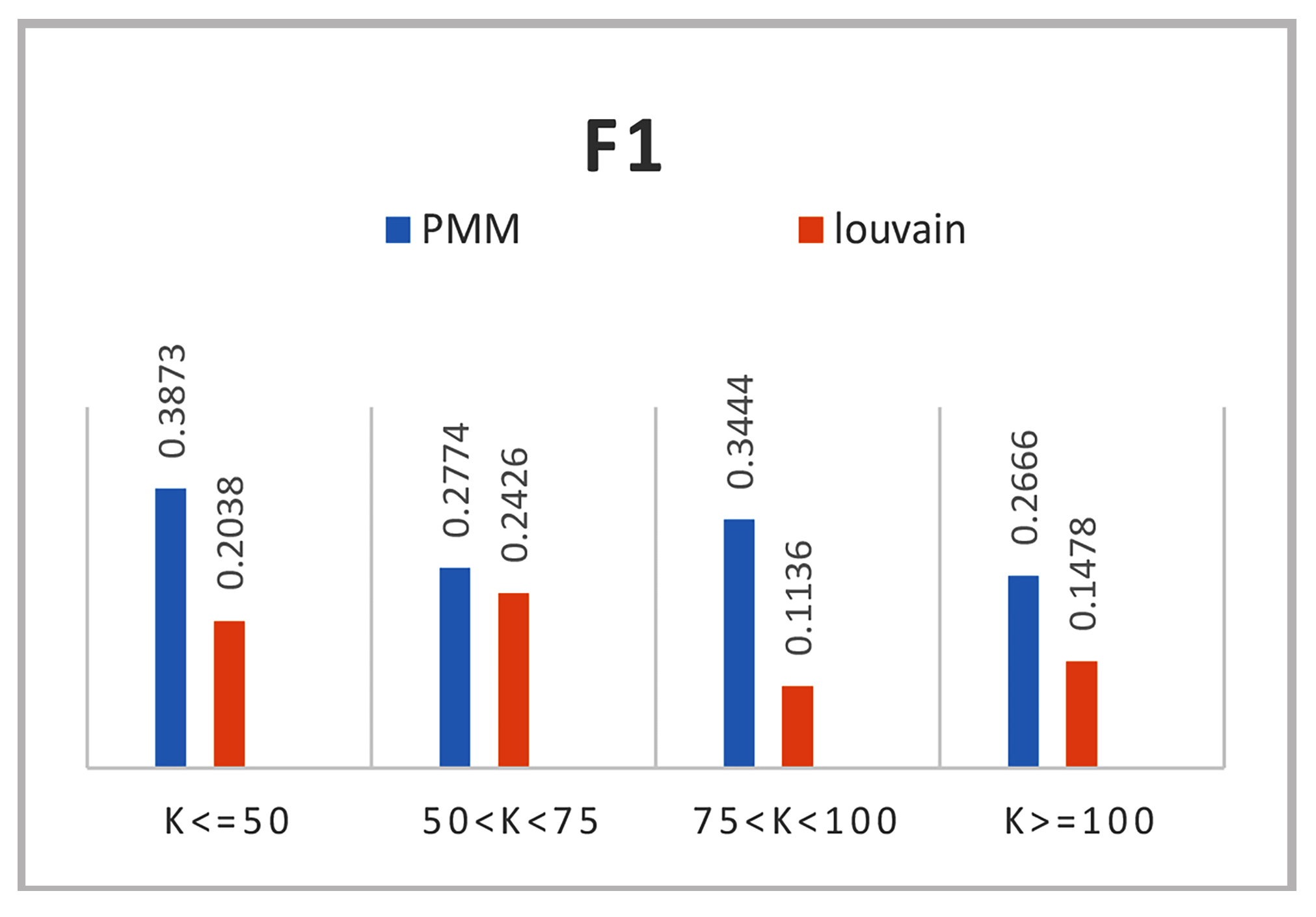}
		\label{f1}
	\end{subfigure}
	\caption{Precision, Recall, and F1 measure for PMM and Louvain.}
	\label{comparison}
\end{figure*}

Having been processed, the output data set is able to easily be used in multi-layer graphs modeling easily. The multi-layer networks can contain structural and content layers. For instance, we have designed a multi-layer network with six layers. The layers consist of the following, followers, co-authorship, skill similarity, equal department, and document similarity. We call this set of data \textbf{M}ultirelation \textbf{R}esearch\textbf{G}ate \textbf{N}network (\textbf{MRGN}). In this data set, each layer is presented as a list-edge data set containing two columns for user $u_i$ and user $u_j$, which shows that there is an edge among user $u_i$ and $u_j$ in the corresponding layer. Following-follower and skill layers' are directed; however, co-authoring and department layers are undirected, but for keeping the consistency of the network, which is directed, we add reciprocal edges. We present the analytical information of each layer in Table \ref{table12}. We utilized multilayer community detection methods for this comparison. After detecting the communities, the users in the same communities are the researchers with higher similarity based on two sources of information, namely, structural and textual information. We use Precision, Recall, and F1 measures for comparing the results, as they have extensively used in the evaluation of recommender systems \cite{shani2011evaluating,kong2016exploiting}. The users who lay in the same communities have higher similarity; therefore, they can be recommended to each other as potential collaborators. We expect the users who lay in the same community have at least a previous collaboration. This collaboration can be each of the six relations indicated in MRGN. For solving this problem, other community detection models \cite{yang2013hierarchical,zhou2018overlapping} can be used. For evaluation, we defined four sets:
\begin{itemize}
	\item recommended and collaborated.
	\item recommended but not collaborated.
	\item not recommended but collaborated. 
	\item not recommended and not collaborated.
\end{itemize}
Then using Precision, Recall, and F1 measure, these sets are compared. Two community detection that we utilized are PMM \cite{jutla2011generalized} and Louvain \cite{tang2009uncovering}. We compared the results for different $K$ communities. The number of communities can be changed for different requirements, as selecting the best $K$ depends on the data and the user's needs. The results are shown in Figure \ref{comparison}. The results show the average of runs on different $K$s. Precision in the PMM is better than Louvain but recall in Louvain is better.

\begin{table*}
	\centering
	\caption{ Information of Multi-relation Networks' Layer.}
	\resizebox{1\textwidth}{!}{%
		\begin{tabular}{ |c | c | c|c |c | c | c| c |}
			\hline	
			Layer &  Number of nodes & Number of edges& Modularity&	Density & Average degree& Diameter & Clustering coefficient\\ \hline
			Cofollovers& 3946 & 39327& 	0.523 & 0.003 & 9.996 & 13 & 0.232\\ \hline
			Cofolloving& 3914& 38383 & 0.531& 	0.003 & 9.807& 13&0.231 \\ \hline
			coAuthors& 1484 & 2579& 0.78 & 	0.002 & 1.738 & 11&0.42 \\ \hline
			Coskills &  3784 & 100000& 0.271&	0.007 & 26.681& 10 & 0.150\\ 
			\hline	
			Cocontent &  3115 & 99962& 0.219&	0.010 & 32.091& 11 & 0.019\\ \hline
			Codepartment &  2970 & 99983& 0.519&	0.011 & 31.271& 10 & 0.025\\ \hline
		\end{tabular}
	}
	\label{table12}
\end{table*}

\subsection{Co-author finding model}
\label{sturctured_data}
Designing a co-author recommender model needs sample-feature prepared data. In the rest of this section, we explain how we processed the raw data to a pair of users table; then, these tables are assessed using a set of well-known classifiers.

\textbf{Preparing pair users table:} The information of this table includes a comparison of pair users among all users using all the extractable information from the social network. This information includes structural and personal information of users. We call this table pair users' comparison table. The rows include a tuple $u_iu_j$. In this table, there are 13 columns which are counter, first user's name, second user's name, followers, followings, answers, living in the same country, living in the same city, h-index measure, RG ranking, department, research similarity, skill similarity, and previous collaboration, respectively. Ten features of these columns can be used for potential future co-authors. The target feature, to be predicted, is previous collaboration.\\

\textbf{Binary features.} Among all features, followers, followings, answers, and living in the same country, living in the same city are binary features and show existence or absence of the features. In addition, previous collaboration, as a target feature, also has a binary value and is 1 if there has been a previous collaboration between the corresponding users.\\

\textbf{Ranking features:} For calculating the values of these features, i.e., h-index and RG ranking, we use Eq. \ref{eq1}. 

\textbf{Textual features:} As mentioned before, for calculating the similarity of textual information, such as skill similarity or department, we have used Elmo.\\

\textbf{A collaborator recommending system modeling:} The processed data includes 10 learning features (independent variables) and one target feature (dependent variable), we call this dataset as \textbf{F}eature \textbf{T}able of \textbf{R}esearch\textbf{G}ate (\textbf{FTRG}). In our data set, there are 13 columns in which the columns stand for first user id, second user id, following, followers, question-answers, country, city, H\_index, RG rank, department, research similarity, skill similarity, and previous collaboration, respectively. The features' value is calculated based on the first and second researchers' raw data. Putting these two columns aside, there remain 10 features that help the predictors to provide their results. Table \ref{table10} demonstrates the results of using our data set in some of the recommending systems. As is clear in Table \ref{table10}, the collaborator recommender methods are evaluated using different classification approaches. Recall, Brier score, AUC, and etc. have been used for assessing the results. Among the classifiers, the decision tree has the best precision (92\%). The results show that the presented methods can be used for designing co-author recommender systems.\\
\\
\begin{table}
	\centering
	\caption{Prediction Models Accuracy}
	\resizebox{1\columnwidth}{!}{%
		\begin{tabular}{ |c|c|c|c|c|c|c|c|c|c|c|c|c| }
			\hline	
			Algorithm & MAE	&RMSE&	TP&	FP&	TN&	FN&	precision&	recall&	f1&	Brier Score &	AUC	& ACC
			\\ \hline	
			AdaBoostM1 & 	0.1933& 	0.2983& 	0.7518& 	0.0155& 	0.9845& 	0.2482& 	0.9768& 	0.7518& 	0.8494& 	0.1318&	0.9293& 	0.8682
			\\ \hline
			Bagging & 	0.1301&	0.2578&	0.8677&	0.0565&	0.9435&	0.1323&	0.9294&	0.8677&	0.8973&	0.0944&	0.9671&	0.9056
			\\ \hline
			BayesNet & 	0.0909&	0.2514&	0.8578&	0.0300&	0.9700&	0.1422&	0.9607&	0.8578&	0.9063&	0.0861&	0.9772&	0.9139
			\\ \hline
			DecisionTable & 	0.1390&	0.2537&	0.8759&	0.0390&	0.9610&	0.1241&	0.9506&	0.8759&	0.9116&	0.0815&	0.9614&	0.9185
			\\ \hline
			LMT&	0.1384&	0.2771&	0.8446&	0.0500&	0.9500&	0.1554&	0.9356&	0.8446&	0.8875&	0.1027&	0.9451&	0.8973
			\\ \hline
			Logistic&	0.1837&	0.3020&	0.7943&	0.0380&	0.9620&	0.2057&	0.9470&	0.7943&	0.8638&	0.1219&	0.9234&	0.8781
			\\ \hline
			MultiClassC	&0.1837&	0.3020&	0.7943&	0.0380&	0.9620&	0.2057&	0.9470&	0.7943&	0.8638&	0.1219&	0.9234&	0.8781
			\\ \hline
			NavieBayesUpdt &	0.1359&	0.3237&	0.8168&	0.0675&	0.9325&	0.1832&	0.9118&	0.8168&	0.8616&	0.1254&	0.9291&	0.8746
			\\ \hline
			RandomCommitee&	0.1203&	0.2758&	0.8818&	0.0835&	0.9165&	0.1182&	0.9004&	0.8818&	0.8909&	0.1009&	0.9489&	0.8991
			\\ \hline
			RandomSubSpace&		0.1946&		0.2659&		0.8545&		0.0310&		0.9690&		0.1455&		0.9593&		0.8545&		0.9037&		0.0882&	0.9748&		0.9118
			\\ \hline
			RandomForect&	0.1259&	0.2587&	0.8698&	0.0633&	0.9368&	0.1302&	0.9217&	0.8698&	0.8949&	0.0967&	0.9652&	0.9033
			\\ \hline
		\end{tabular}
	}
	\label{table10}
\end{table}

\section{Conclusion}
\label{Conclusion}
In this research, we tried to prepare and provide a processed data set to be used in collaborator finding models. To do so, at first, we extracted the raw data from ResearchGate social networks. ResearchGate is one of the most popular academic social networks which have attracted a huge amount of researchers' attention in the past few years. These academic social networks provide the most information compared to other public social networks. Afterward, we applied a set of processes and analyses on the raw data. These processes are so important because, without them, the data set could not be prepared and used in collaborator systems. For instance, raw data are not able to provide a sufficient amount of information for recommender systems. One of the methods important features for detecting similarity of the researchers is comparing their research outputs. In the preprocessing phase, we extracted the content similarity of users using a state-of-the-art deep learning method. By doing so, we extracted the content similarity of the users.\\
After preparing and analyzing data, we eventually prepared two datasets from raw data. One of these models includes six layers, each of which indicates a special aspect of users' relations. The second model is a pairwise table that each cell of a matrix denotes the similarity of two corresponding users. These two tables are evaluated separately. One of the models is the collaborator finding model, which uses a multilayer network, and the next is the co-author finding model, which uses a pair of user tables. The first model is a collaborator recommender model, which has been evaluated using some well-known multilayer algorithms. The second model has been assessed using a number of classification methods. To the best of our knowledge, there is no processed and analyzed dataset with this size.\\
In fact, the information directionality of our dataset makes it suitable for the researches which needs graphs for analyzing data, specifically for the ones which need to have direct graphs in their analysis. It worth mentioning that the preprocess of data is done in this line. Furthermore, the raw data of our data set can be used in other research fields, such as scientific text analysis, expert finding, communication, and so forth. In turn, having been processed and adopted, the data can be used for evaluating algorithms in different branches of science. Using the published dataset, future research has access to an enriched set of the dataset for evaluating their works.
\bibliographystyle{IEEEannot}  

\bibliography{references}

\end{document}